\documentclass[5p,sort&compress,times]{elsarticle}
\usepackage{lineno,hyperref}
\journal{arXiv}
\usepackage[T1]{fontenc}
\hypersetup{colorlinks=true,allcolors=[rgb]{0,0,0.6}}
\usepackage{amsmath,amsfonts,amssymb,amsthm}
\usepackage{braket,mleftright}
\usepackage{graphicx,caption,subcaption}
\newcommand{\R}{\mathbb{R}}
\newcommand{\N}{\mathbb{N}}
\newcommand{\mrm}[1]{\mathrm{#1}}
\newcommand{\ol}[1]{\overline{#1}}
\newcommand{\abs}[1]{\lvert#1\rvert}
\newcommand{\img}{\mrm{i}}
\newcommand{\e}{\mrm{e}}
\newcommand{\ie}{\textit{i.e.}\;}
\newcommand{\eg}{\textit{e.g.}\;}
\newcommand{\cgc}[6]{\begin{bmatrix}
#1 & #2 & #3 \\ #4 & #5 & #6\end{bmatrix}}
\newcommand{\dd}{\,\mrm{d}}
\DeclareMathOperator{\dv}{div}
\DeclareMathOperator{\grad}{grad}
\numberwithin{equation}{section}
\allowdisplaybreaks
\begin{document}
\begin{frontmatter}
\title{Irreducible tensor form for the AME coupling}
\author{Rytis Jur\v{s}\.{e}nas}
\address{Vilnius University, Center for Physical 
Sciences and Technology, Institute of Theoretical
Physics and Astronomy, Saul\.{e}tekio ave. 3,
Vilnius 10222, Lithuania}
\ead{rytis.jursenas@tfai.vu.lt}
\begin{abstract}
Using the multipole expansion of electromagnetic (EM) field,
we present the angular magnetoelectric (AME) coupling in 
irreducible tensor form. We evaluate the matrix elements when 
the radiation source is described by electronic transitions in
atomic systems. The results indicate that the energy corrections 
increase for short wavelengths and large charge number.
\end{abstract}
\end{frontmatter}
\section{Introduction}
In consequence of their results on the time-dependent Foldy--Wouthuysen 
transformation \cite{Mondal2}, Mondal et al show \cite{Mondal} that 
the AME coupling represents the $O(c^{-2})$ correction to the Dirac 
operator for an electron in a potential $V$ and interacting with the 
EM field, which is characterized by the $4$-component vector potential 
$A^\mu=(\Phi,A)$ (for an alternative derivation 
we refer to \cite{OConnell}):
\begin{equation}
H_{AME}=\frac{1}{2c^2}(H^{in}_{AME}+H^{ex}_{AME})
\label{eq:AME}
\end{equation}
where 
\begin{subequations}\label{eq:in-ex}
\begin{align}
H^{in}_{AME}=&
S\cdot(E^{in}\times A)\,, \quad 
E^{in}=-\grad V\,, 
\\ 
H^{ex}_{AME}=&S\cdot(E^{ex}\times A)\,, \quad 
E^{ex}=-\dot{A}-\grad\Phi
\end{align}
\end{subequations}
and $S$ is the spin operator.
Throughout we use atomic units unless explicitly stated otherwise. One 
calls $H^{in}_{AME}$ (resp. $H^{ex}_{AME}$) an intrinsic 
(resp. induced) part of AME coupling. In the Coulomb gauge fixing, 
$H^{ex}_{AME}$ serves as the source for obtaining the "hidden energy" 
that couples the EM angular momentum density with magnetic moments 
\cite{Rael}.

It is our purpose here to investigate the input of 
$H_{AME}$ into the atomic energy levels provided that the radiation 
source is defined by electronic transitions. The exploration of the standard
multipole expansion of EM field \cite{Blatt} allows us to represent 
$H_{AME}$ as the summable series (in the sense of distributions) of 
irreducible tensor operators, while the information about the  
radiation source is contained separately, in the coefficients of expansion 
(amplitudes). 

As one could expect, the contribution of the AME coupling to the 
total energy should be small enough. For example, we show that 
the matrix elements of the intrinsic part are of order 
$O(\omega^2Z^2c^{-3})$ for the $E1$ transition, while the matrix 
elements of the induced part are of order $O(\omega^5Z^{-2}c^{-6})$ 
for the same type of radiation; here $\omega$ is 
the transition energy. However, $\omega$ usually increases as the 
charge number $Z$ becomes large, which results in larger energy 
corrections. 

By \eqref{eq:in-ex}, $H^{in}_{AME}$ is 
time-dependent, while $H^{ex}_{AME}$ is 
time-independent. The latter is in agreement with \cite{Mondal}, where 
the plane EM wave expansion is used for explaining the inverse Faraday 
effect. In addition, we show that $H^{ex}_{AME}$ can be split in 
two separate parts. One part is traditional in the sense that 
it does not vanish if the multipole moment of order $l\in\N$ 
is nonzero for at least one fixed $l$, 
while the second one is more "exotic" in the sense that 
it can be nonzero only if the multipole 
moment is nonzero for at least two different $l$. The latter 
case arises when, for example, one considers electronic satellite 
transitions produced by electron capture and subsequent radiative decay 
\cite{Bei14,Safr12,Safr11,Chab94-1,Chab94-2}.

In Sec.~\ref{sec:tensor} we express \eqref{eq:in-ex} in 
irreducible tensor form. We work in the Coulomb gauge fixing and 
we use the standard technique of angular momentum theory 
\cite{Rudz2,Rudz,Jucys-1,Jucys-2} (including the notation and 
the phase system used therein). We discuss the matrix elements 
in particular cases in Sec.~\ref{sec:matrix}.
\section{Tensor operators}\label{sec:tensor}
\subsection{Amplitudes}
Let the radiation of energy 
$\omega=\nu c=E_{\alpha J}-E_{\alpha^\prime J^\prime}$ 
be emitted by the electron going 
from the state $\ket{\alpha JM}$ to the (lower) state 
$\ket{\alpha^\prime J^\prime M^\prime}$; $\alpha$ and $\alpha^\prime$
denote additional quantum numbers if necessary. When 
$\nu\ll1$, the amplitudes for the radiation  
of order $(l,m)$, $l\in\N$, $m\in\{-l,\ldots,l\}$, are approximated by
\cite{Blatt} 
\begin{equation}
a^\#_{\nu l m}=
\delta_{m\rho}a^\#_{\nu l}\,, \quad 
\rho=M-M^\prime\,.
\label{eq:a}
\end{equation}
Here the superscript denotes both $E$ (electric type) and 
$M$ (magnetic type), and 
\begin{equation}
a^E_{\nu l}=\lambda_{\nu l}Q_{\nu l}\,, \quad 
a^M_{\nu l}=-\lambda_{\nu l}M_{\nu l}\,. 
\label{eq:aEM1}
\end{equation}
The multiplier 
\begin{align}
\lambda_{\nu l}=&(-1)^{J-J^\prime+1}
\frac{\img^{-l}\nu^{l+2}K_l}{(2l+1)!!}
\frac{\sqrt{4\pi(2l+1)}}{\sqrt{2J+1}} 
\nonumber \\ 
&\cdot 
\cgc{J^\prime}{l}{J}{M^\prime}{\rho}{M}\,, \quad 
K_l=-\sqrt{1+1/l}\,.
\label{eq:lambda}
\end{align}
The number $Q_{\nu l}$ (resp. $M_{\nu l}$) is the reduced matrix 
element of the electric (resp. magnetic) 
multipole moment $Q^l$ (resp. $M^l$):
\begin{equation}
Q_{\nu l}=(\alpha^\prime J^\prime\Vert Q^l\Vert 
\alpha J)\,, \quad 
M_{\nu l}=(\alpha^\prime J^\prime\Vert M^l\Vert 
\alpha J)\,.
\end{equation}
When the magnetization is ignored, we have 
\begin{equation}
Q^l=-r^lC^l\,, \quad 
M^l=-\frac{\sqrt{l(2l-1)}}{c(l+1)}
r^{l-1}[C^{l-1}\times L^1]^l\,. 
\label{eq:QlMl}
\end{equation}
Otherwise: $Q^l$ is replaced by $Q^l+O(\nu/c)$, hence 
we omit the $O(\nu/c)$ correction since we already 
have the small $\nu^{l+2}$ in \eqref{eq:lambda};
$M^l$ is replaced by $M^l+M^{\prime\,l}$, where 
\begin{equation}
M^{\prime\,l}=-\frac{1}{c}\sqrt{l(2l-1)}
r^{l-1}[C^{l-1}\times S^1]^l\,.
\end{equation}
In the examples to be followed, we assume 
$M^l+M^{\prime\,l}$ when we write $M^l$; see also 
\cite[Secs.~4 and 25]{Rudz2}.
\subsection{Intrinsic part}\label{sec:in}
As in \cite{Mondal}, we take the real part of the external 
electric field $E^{ex}$. Applying the well-known 
angular momentum technique, we deduce from 
\cite[Appendix~B.2]{Blatt} the following form for the intrinsic 
part $H^{in}_{AME}\equiv (H^{in}_{AME})^E_{\nu t}$ of electric type 
\begin{subequations}
\begin{equation}
(H^{in}_{AME})^E_{\nu t}=
\sum_{l\in\N}\sum_{m=-l}^l 
\alpha^E_{\nu l m}(t) (H^{in}_{AME})^E_{\nu l m} 
\label{eq:HinAME-E-0}
\end{equation}
where the rank-$l$ irreducible tensor operator
\begin{align}
(H^{in}_{AME})^E_{\nu l}=&
\frac{\img^{-l}V^\prime(r)}{2\omega\sqrt{\pi(2l+1)}}
[C^l\times S^1]^l
\nonumber \\ 
&\cdot
[(l+1)j_{l-1}(\nu r)-lj_{l+1}(\nu r)] 
\label{eq:HinAME-E-1}
\end{align}
($j_l$ is the spherical Bessel function)
and the amplitude
\begin{equation}
\alpha^E_{\nu l m}(t)=
\frac{1}{2}\bigl(\e^{-\img(\omega t+\sigma_l)}a^E_{\nu l m}
+(-1)^m\e^{\img(\omega t+\sigma_l)}\ol{a^E_{\nu l,-m}} \bigr)\,. 
\label{eq:alphaE}
\end{equation}
\end{subequations}
Here $\sigma_l=\arg\Gamma(l+1+\img\eta)$ is the Coulomb phase 
shift, $\eta=-Z/\nu$ is the Sommerfeld parameter. 

Likewise, the intrinsic 
part $H^{in}_{AME}\equiv (H^{in}_{AME})^M_{\nu t}$ of magnetic type 
is written in the form \eqref{eq:HinAME-E-0}, but with the superscript $E$ 
replaced by the superscript $M$, and with the amplitude replaced by 
\begin{subequations}
\begin{equation}
\beta^M_{\nu l m}(t)=
\frac{1}{2}\bigl(\e^{-\img(\omega t+\sigma_l)}a^M_{\nu l m}
-(-1)^m\e^{\img(\omega t+\sigma_l)}\ol{a^M_{\nu l,-m}} \bigr)\,.
\label{eq:betaM}
\end{equation}
The corresponding rank-$l$ irreducible tensor operator
\begin{align}
(H^{in}_{AME})^M_{\nu l}=&
\frac{\img^{-l-1}V^\prime(r)j_l(\nu r)}{2\omega\sqrt{\pi(2l+1)}}
\nonumber \\ 
&\cdot
\bigl(\sqrt{(l+1)(2l-1)}[C^{l-1}\times S^1]^l 
\nonumber \\ 
&-\sqrt{l(2l+3)}[C^{l+1}\times S^1]^l \bigr)\,. 
\label{eq:hin-M1} 
\end{align}
\end{subequations}

In \cite{Mondal} the authors put   
$A=B\times x/2$ for almost every $x\in\R^3$. In this case $\dv A^M=0$ but 
$\dv A^E\neq0$; for $A=A^M$ of magnetic type, $j_l(\nu r)$ 
in \eqref{eq:hin-M1} is replaced by 
$[(l+2)j_l(\nu r)-\nu r j_{l+1}(\nu r)]/2$.

From the point of view of energy levels, 
the treatment of $H^{in}_{AME}/(2c^2)$, when considered as the $O(c^{-2})$ 
correction to the Pauli operator for an electron in a potential 
$V$, is subtle in that it is time-dependent. We refer to 
\cite{Bou08,Mau99,Lew69}, 
where the eigenvalue problem for the time-dependent Pauli equation 
is studied in detail.
\subsection{Induced part}\label{sec:ind}
Unlike the intrinsic part of AME coupling, the induced part 
contains the products of the time-dependent amplitudes 
$\alpha^\#_{\nu l m}(t)$ and $\beta^\#_{\nu l^\prime m^\prime}(t)$;
here $\alpha^M_{\nu l m}(t)$ (resp. $\beta^E_{\nu l m}(t)$) is 
defined by \eqref{eq:alphaE} (resp. \eqref{eq:betaM}), but with the 
superscript $E$ (resp. $M$) replaced by the superscript $M$ 
(resp. $E$). However, using the symmetry properties of the 
products and interchanging the summation indices $l$ and 
$l^\prime$ we find that 
the induced part of AME coupling is actually time-independent. 
As a result, $H^{ex}_{AME}\equiv (H^{ex}_{AME})^\#_\nu$ splits 
into two parts:
\begin{equation}
(H^{ex}_{AME})^\#_\nu=(H^{ex}_{AME})^{\#\,\prime}_\nu
+(H^{ex}_{AME})^{\#\,\prime\prime}_\nu\,.
\end{equation} 
For the radiation of electric type we have 
\begin{subequations}
\begin{align}
(H^{ex}_{AME})^{E\,\prime}_\nu=&
\frac{(-1)^\rho}{2}\sum_l
\abs{a^E_{\nu l}}^2 
\nonumber \\ 
&\cdot 
\sum_{J=\text{odd}}
(H^{ex}_{AME})^E_{\nu l l J0}
\cgc{l}{l}{J}{\rho}{-\rho}{0}\,, 
\label{eq:ex-Ep} \\ 
(H^{ex}_{AME})^{E\,\prime\prime}_\nu=&
(-1)^\rho\sum_{l<l^\prime}
\sum_J \gamma^E_{\nu l l^\prime J}
(H^{ex}_{AME})^E_{\nu l l^\prime J0} 
\nonumber \\ 
&\cdot
\cgc{l}{l^\prime}{J}{\rho}{-\rho}{0} 
\label{eq:ex-Epp}
\end{align}
\end{subequations}
with $l\geq\max\{1,\abs{\rho}\}$. The "amplitude"
\begin{equation}
\gamma^E_{\nu l l^\prime J}=
\frac{1}{2}\bigl(
\e^{-\img(\sigma_l-\sigma_{l^\prime})}
a^E_{\nu l}\ol{a^E_{\nu l^\prime}}
-(-1)^{l+l^\prime+J}
\e^{\img(\sigma_l-\sigma_{l^\prime})}
\ol{a^E_{\nu l}}a^E_{\nu l^\prime} \bigr) 
\label{eq:gammaE}
\end{equation}
and $(H^{ex}_{AME})^E_{\nu l l^\prime J0}$ is the 
$0$th component of the rank-$J$ ($\abs{l-l^\prime}\leq J \leq 
l+l^\prime$) tensor operator 
\begin{align}
&(H^{ex}_{AME})^E_{\nu l l^\prime J}=
\frac{\sqrt{3/2}\,\img}{2\pi\omega}(-1)^{J+1}
\sum_k\img^k\sqrt{2k+1}[C^k\times S^1]^J  
\nonumber \\ 
&\cdot 
\Bigl(
\sqrt{l(2l+3)l^\prime(2l^\prime+3)}
j_{l+1}(\nu r)j_{l^\prime+1}(\nu r) 
\nonumber \\ 
&\cdot 
\cgc{l+1}{l^\prime+1}{k}{0}{0}{0}
\begin{Bmatrix}
l & 1 & l+1 \\ l^\prime & 1 & l^\prime+1 \\ J & 1 & k
\end{Bmatrix} 
\nonumber \\ 
&-
\sqrt{l(2l+3)(l^\prime+1)(2l^\prime-1)}
j_{l+1}(\nu r)j_{l^\prime-1}(\nu r) 
\nonumber \\ 
&\cdot 
\cgc{l+1}{l^\prime-1}{k}{0}{0}{0}
\begin{Bmatrix}
l & 1 & l+1 \\ l^\prime & 1 & l^\prime-1 \\ J & 1 & k
\end{Bmatrix} 
\nonumber \\ 
&-
\sqrt{(l+1)(2l-1)l^\prime(2l^\prime+3)}
j_{l-1}(\nu r)j_{l^\prime+1}(\nu r) 
\nonumber \\ 
&\cdot 
\cgc{l-1}{l^\prime+1}{k}{0}{0}{0}
\begin{Bmatrix}
l & 1 & l-1 \\ l^\prime & 1 & l^\prime+1 \\ J & 1 & k
\end{Bmatrix} 
\nonumber \\ 
&+
\sqrt{(l+1)(2l-1)(l^\prime+1)(2l^\prime-1)}
j_{l-1}(\nu r)j_{l^\prime-1}(\nu r) 
\nonumber \\ 
&\cdot 
\cgc{l-1}{l^\prime-1}{k}{0}{0}{0}
\begin{Bmatrix}
l & 1 & l-1 \\ l^\prime & 1 & l^\prime-1 \\ J & 1 & k
\end{Bmatrix} \Bigr)\,. 
\label{eq:HexAME-E-2}
\end{align}
The integers $k$ are such that 
$k+l+l^\prime$ is even and at least one of the following four 
conditions holds: 
\begin{itemize}
\item[$\circ$] 
$\max\{\abs{J-1},\abs{l-l^\prime}\}\leq k\leq 
\min\{J+1,l+l^\prime\pm2\}$
\item[$\circ$] 
$\max\{\abs{J-1},\abs{l-l^\prime\pm2}\}\leq k\leq 
\min\{J+1,l+l^\prime\}$. 
\end{itemize}
Notice that $J$ in \eqref{eq:ex-Ep} is necessarily odd, because 
\begin{equation}
(H^{ex}_{AME})^E_{\nu l^\prime l J}=
(-1)^{l+l^\prime+J+1}
(H^{ex}_{AME})^E_{\nu l l^\prime J}\,.
\label{eq:symm}
\end{equation}

It follows from above that:
\begin{itemize}
\item[$-$] 
$(H^{ex}_{AME})^{E\,\prime}_\nu=0$ for $\rho=0$
(\ie $M=M^\prime$).
\item[$-$] 
$(H^{ex}_{AME})^{E\,\prime\prime}_\nu$ is nonzero if 
$Q_{\nu l}$ is nonzero for at least two different values of $l$.
\end{itemize} 

For the radiation of magnetic type, the terms 
$(H^{ex}_{AME})^{M\,\prime}_\nu$
and $(H^{ex}_{AME})^{M\,\prime\prime}_\nu$ are given by 
\eqref{eq:ex-Ep} and \eqref{eq:ex-Epp}, respectively, but with 
the superscript $E$ replaced by the superscript $M$, the phase 
$(-1)^\rho$ replaced by $(-1)^{\rho+1}$, and with the 
corresponding rank-$J$ tensor operator 
\begin{align}
(H^{ex}_{AME})^M_{\nu l l^\prime J}=&
\frac{\sqrt{3/2}\,\img}{2\pi\omega}(-1)^{J+1} 
(2l+1)(2l^\prime+1) 
\nonumber \\ 
&\cdot 
j_l(\nu r) j_{l^\prime}(\nu r)
\sum_k\img^k\sqrt{2k+1}[C^k\times S^1]^J
\nonumber \\ 
&\cdot 
\cgc{l}{l^\prime}{k}{0}{0}{0} 
\begin{Bmatrix}
l & 1 & l \\ l^\prime & 1 & l^\prime \\ J & 1 & k
\end{Bmatrix}\,. 
\label{eq:HexAME-M-2}
\end{align}
The integers $k$ are such that $k+l+l^\prime$ is even and it holds 
$\max\{\abs{J-1},\abs{l-l^\prime}\}\leq k \leq
\min\{J+1,l+l^\prime\}$.
For $A=B\times x/2$ of magnetic type, $j_{l^\prime}(\nu r)$ in 
\eqref{eq:HexAME-M-2} is replaced by $[(l^\prime+2)j_{l^\prime}(\nu r)-
\nu r j_{l^\prime+1}(\nu r)]/2$. 

Notice that $(H^{ex}_{AME})^M_{\nu l l^\prime J}$ satisfies the 
symmetry property analogous to \eqref{eq:symm}. 
In addition:
\begin{itemize}
\item[$-$] 
Since the $9j$-symbol   
$\Bigl\{\begin{smallmatrix}
l & 1 & l \\ l^\prime & 1 & l^\prime \\ J & 1 & k
\end{smallmatrix}\Bigr\}=0$
for $k=J$ (see \eg \cite[Eq.~(31.10)]{Jucys-1}), and 
since $k+l+l^\prime$ is even, the total 
$(H^{ex}_{AME})^M_\nu=0$ for $\rho=0$.
\item[$-$] 
$(H^{ex}_{AME})^{M\,\prime\prime}_\nu$ is nonzero if 
$M_{\nu l}$ is nonzero for at least two different values of $l$
(provided $\rho\neq0$).
\end{itemize}

Although it should be obvious, we would like to emphasize that 
$J$ in Sec.~\ref{sec:ind} is not the same as 
$J$ in \eqref{eq:a}.
\section{Matrix elements}\label{sec:matrix}
The irreducible tensor form of the AME coupling is convenient in 
that one can directly apply the Wigner--Eckart theorem, which we 
take here of the form \cite[Eq.~(5.15)]{Rudz2}. Therefore, we do 
not rewrite the obtained tensor operators in matrix form, but rather 
concentrate on some particular cases of electronic transitions.

We use the standard non-relativistic basis 
functions \cite[Eq.~(2.16)]{Rudz2} $\ket{nlj\mu}$ enumerated by rationals  
$n\in\N$, $0\leq l\leq n-1$, $\abs{l-1/2}\leq j\leq l+1/2$, and 
$-j\leq \mu\leq j$. Then, the reduced matrix elements of type 
$(l_1j_1\Vert [C^k\times S^1]^J\Vert l_2j_2)$ are found 
by using \eg \cite[Eq.~(38.21)]{Jucys-1}. Notice that 
$H^{ex}_{AME}$ is diagonal in $\mu$.

The radial integrals that we need are of the form 
\begin{subequations}
\begin{align}
\braket{r^l}_{\alpha_1,\alpha_2}=&
\int_0^\infty\ol{R_{\alpha_1}(r)}
R_{\alpha_2}(r)r^{l+2}\dd r\,, 
\label{eq:rad-1} \\ 
I_{\nu l}(\alpha_1,\alpha_2)=&
\int_0^\infty\ol{R_{\alpha_1}(r)}V^\prime(r)
j_l(\nu r)R_{\alpha_2}(r)r^2\dd r\,, 
\label{eq:rad-2} \\ 
I_{\nu ll^\prime}(\alpha_1,\alpha_2)=&
\int_0^\infty\ol{R_{\alpha_1}(r)}j_l(\nu r)
j_{l^\prime}(\nu r)R_{\alpha_2}(r)r^2\dd r\,.
\label{eq:rad-3}
\end{align}
\end{subequations}
The integral \eqref{eq:rad-1} is used for the calculation of  
reduced matrix elements $Q_{\nu l}$ and $M_{\nu l}$. 
The integral \eqref{eq:rad-2} 
(resp. \eqref{eq:rad-3}) appears in the matrix elements of 
the intrinsic (resp. induced) part of the AME coupling. 

Thus, given the transition of order $l$, one finds that 
the intrinsic part of electric (resp. magnetic) 
type is of order $O(\nu^{2l})$ (resp. $O(\nu^{2l+1})$); 
the induced part $(H^{ex}_{AME})^{E\,\prime}_\nu$ 
(resp. $(H^{ex}_{AME})^{M\,\prime}_\nu$) is of order 
$O(\nu^{4l+1})$ (resp. $O(\nu^{4l+3})$). In particular, 
it follows that 
the energy corrections of both $(H^{in}_{AME})^M_\nu$ for 
$M2$ transition and $(H^{ex}_{AME})^{E\,\prime}_\nu$ for 
$E1$ transition are of order $O(\nu^5)$.

For hydrogenic radial functions we have $\alpha=(n,l)$.
The radial functions $R_{nl}$ are sufficient 
to evaluate the order of magnitude of matrix elements. Moreover, 
these functions are convenient for analytical purposes.

Since the AME coupling represents the $O(c^{-2})$ correction, the large 
Dirac component of the relativistic wave function or the large radial 
component calculated from non-relativistic MCHF radial functions 
\cite[Eq.~(131)]{G0} would fit in this approach as well. The 
calculations based on these functions are left for future 
investigations. 

From now on we include in the expressions of tensor operators 
the previously excluded multiplier $(2c^2)^{-1}$ (recall \eqref{eq:AME}).
\subsection{Multipole radiation of a fixed order}
Let us consider the transition 
$2s2p^2\,^2\!P_{1/2}-2s^22p\,^2\!P^o_{3/2}$. 
Using \cite[Eqs.~(25.28) and (27.7)]{Rudz2} we get that
\[Q_{\nu l}=\delta_{l1}3\sqrt{2}/Z\,, \quad 
M_{\nu l}=\delta_{l2}5/(4\sqrt{3}cZ)\,.\]

For $V(r)=-Z/r$ and $M=M^\prime=1/2$ (\ie $\rho=0$), the first nonzero 
matrix elements in the intrinsic case are given by 
\begin{subequations}\label{eq:eddss00}
\begin{align}
&\braket{1s,\pm1/2\vert 
(H^{in}_{AME})^E_{\nu t} \vert 
2p_-,\pm1/2} 
\\ 
&\quad 
=\braket{2p_-,\pm1/2\vert 
(H^{in}_{AME})^E_{\nu t} \vert 
1s,\pm1/2} 
\\ 
&\quad 
=\mp\frac{4\sqrt{2}}{162}
\frac{Z^2\nu^2}{c^3} 
\sin(\omega t+\sigma_1)+O(\nu^4)\,, 
\nonumber \\ 
&\braket{1s,\pm1/2\vert 
(H^{in}_{AME})^M_{\nu t} \vert 
2p_+,\pm1/2} 
\nonumber \\ 
&\quad 
=\ol{\braket{2p_+,\pm1/2\vert 
(H^{in}_{AME})^M_{\nu t} \vert 
1s,\pm1/2}} 
\nonumber \\ 
&\quad 
=\mp\frac{\img}{1215\sqrt{10}}
\frac{\nu^5}{c^4} 
\sin(\omega t+\sigma_2)+O(\nu^7)\,. \nonumber 
\end{align}
\end{subequations}
According to \cite[Tab.~3]{G}, $\omega=\omega(Z)$ 
is an increasing function; for example, $\omega$ is around $597402$ 
(cm$^{-1}$) in Ca XVI and around $847413$ (cm$^{-1}$) in 
Zn XXVI. This indicates that the contribution of the intrinsic part of 
AME coupling to the total energy increases when $Z$ becomes large.

\begin{figure}[htp!]
\centering
\includegraphics[width=.42\textwidth]{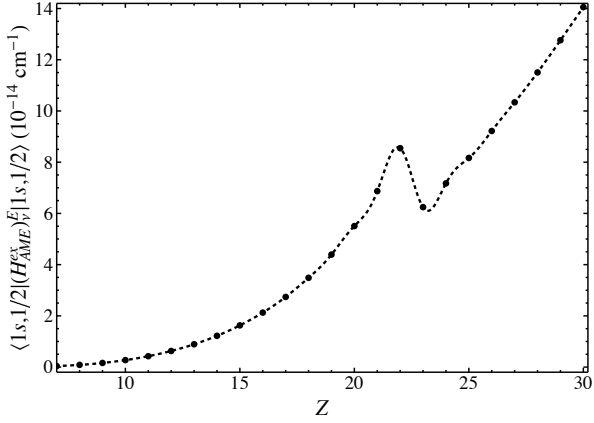}
\caption{$\braket{1s,1/2\vert 
(H^{ex}_{AME})^E_\nu \vert 
1s,1/2}$ as a function of $Z$;
the transition energies $\omega=\omega(Z)$ for $Z=7$ to $30$
are adapted from \cite[Tab.~3]{G}. The dotted curve represents
the degree four polynomial that approximates data points. 
The kink at around $Z=23$ corresponds to 
$\omega(Z=23)<\omega(Z=22)$.}
\label{fig:F1}
\end{figure}
For $M=-M^\prime=1/2$ (\ie $\rho=1$), the first nonzero 
matrix elements in the induced case are given by 
(note that the tensor operator in \eqref{eq:ex-Epp} vanishes)
\begin{align}
\braket{1s,\pm1/2\vert 
(H^{ex}_{AME})^E_\nu \vert 
1s,\pm1/2} 
=&\pm\frac{1}{12}
\frac{\nu^5}{c^3Z^2}
+O(\nu^7)\,, 
\label{eq:eddss0} \\ 
\braket{1s,\pm1/2\vert 
(H^{ex}_{AME})^M_\nu \vert 
1s,\pm1/2} 
=&\pm\frac{1}{184320}
\frac{\nu^{11}}{c^3Z^6}
+O(\nu^{13})\,. \nonumber 
\end{align}
We plot $\nu^5/(12c^3Z^2)$ in Fig.~\ref{fig:F1}, from where 
we conclude that the increasing $\nu^5$ tends to dominate over 
$Z^{-2}$.

\begin{table}[htp!]
\centering
\caption{\label{tab:tab1}The value of $\braket{1s,1/2\vert 
(H^{ex}_{AME})^E_\nu \vert 
1s,1/2}$ for a given electronic transition in Zn XXVI. The 
numbers are in $10^{-13}$cm$^{-1}$.}
\begin{tabular}{@{}lr}
\hline
Transition & Matrix element \\
\hline
$2s2p^2\,^2\!S_{1/2}-2s^22p\,^2\!P^o_{1/2}$ & $16.3051$ \\ 
$2s2p^2\,^2\!S_{1/2}-2s^22p\,^2\!P^o_{3/2}$ & $3.2119$ \\ 
$2s2p^2\,^2\!P_{1/2}-2s^22p\,^2\!P^o_{1/2}$ & $35.9883$ \\ 
$2s2p^2\,^2\!P_{3/2}-2s^22p\,^2\!P^o_{3/2}$ & $40.8814$ \\
$2s2p^2\,^2\!D_{3/2}-2s^22p\,^2\!P^o_{1/2}$ & $4.4750$ \\ 
$2s2p^2\,^2\!D_{5/2}-2s^22p\,^2\!P^o_{3/2}$ & $2.0247$ \\ 
\hline
\end{tabular}
\end{table}
The matrix elements for some $E1$ transitions are listed in 
Tab.~\ref{tab:tab1}. In comparison, for $E2$ ($M1$) transition 
$2s^22p\,^2\!P^o_{3/2}-2s^22p\,^2\!P^o_{1/2}$ the matrix element of 
$H^{ex}_{AME}$  
is of order $O(\nu^9Z^{-6}c^{-3})$ ($O(\nu^{11}Z^{-2}c^{-3})$).
\subsection{Electronic satellite transitions}
So far we have considered transitions of a particular order $l$.
Now we give examples for the term in \eqref{eq:ex-Epp}. 

Let the radiation source be the decay of type 
\[n_1l_1^{4l_1+2}n_2l_2^{4l_2+2}n_3l_3n_4l_4-
n_1l_1^{4l_1+2}n_2l_2^{4l_2+2}n_3^\prime l_3^\prime n_4l_4\]
of wavelength $\gg21$ ($\textrm{\AA}$), \ie $\nu\ll1$. The  
radiation of this type produces satellite lines considered 
\eg in \cite{Safr12}.
For the transition in Yb-like W \cite[Tab.~6]{Safr12}
\[4f^{14} 5p^6 5f 6d \,^3\!H_6-4f^{14} 5p^6 5d 6d \,^3\!G_5\]
of wavelength $410.6$ ($\textrm{\AA}$) (that is, 
$\nu\approx0.050928$) we get that $Q_{\nu l}$ is nonzero for 
$l=1,3,5$, where 
\begin{align*}
Q_{\nu 1}=&\sqrt{\frac{39}{7}}
\braket{r^1}_{5d,5f}\,, \quad 
Q_{\nu 3}=\frac{1}{3}\sqrt{\frac{26}{3}}
\braket{r^3}_{5d,5f}\,, 
\\ 
Q_{\nu 5}=&\frac{1}{33}\sqrt{\frac{2210}{21}}
\braket{r^5}_{5d,5f}
\end{align*}
and where the radial integrals 
\begin{align*}
\braket{r^l}_{5d,5f}=&
-\frac{1}{63}2^{-l-7}5^{l-2}Z^{-l} 
\\ 
&\cdot 
(l+2)[16+l(l+3)]\Gamma(l+8)\,.
\end{align*}
Let $M=1$ and $M^\prime=0$ so that $\rho=1$. The matrix 
elements of $(H^{ex}_{AME})^{E\,\prime\prime}_\nu$ are of 
order $O(\nu^9Z^{-6}c^{-3})$; \eg 
\begin{subequations}\label{eq:eddss}
\begin{align}
&\braket{1s,1/2\vert (H^{ex}_{AME})^{E\,\prime\prime}_\nu \vert 
3d_-,1/2} 
\\ 
&\quad 
=\ol{\braket{3d_-,1/2\vert (H^{ex}_{AME})^{E\,\prime\prime}_\nu \vert 
1s,1/2}} 
\\ 
&\quad 
=\frac{3825\sqrt{15}\,\img}{2464}
\frac{\nu^9}{c^3Z^6}\sin(\sigma_1-\sigma_3) 
+O(\nu^{11})\,. \nonumber
\end{align}
\end{subequations}
Therefore, the eigenvalues of the (Hermitian) matrix
are of order $10^{-24}\sin(\sigma_1-\sigma_3)$ (cm$^{-1}$). 
To compare with, for $M^\prime=-1$ the above 
matrix element is of order $O(\nu^{17}Z^{-14}c^{-3})
\sin(\sigma_3-\sigma_5)$. 

There are transitions (\eg when $l=2,4$) for which the matrix 
elements are proportional to $\cos(\sigma_l-\sigma_{l^\prime})$; hence 
theoretically the elements are nonzero even if one puts the Coulomb phase 
shift to $0$. On the other hand, the above example suggests that 
the contribution of the present part of AME coupling is practically 
negligible.

The major cause of the small values of matrix elements 
is the small $\nu$. To make $\nu$ large the satellite lines are 
a good example, for their wavelengths can be less than 
$21$ ($\textrm{\AA}$) \cite{Bei14,Safr11}. In order to use  
the transitions with $\nu\geq1$ we need to modify the definition of
amplitudes. Here we only consider the radiation of electric type.
With the help of \cite[Appendix~B, Eq.~(4.10)]{Blatt} we find that 
the multiplier $\lambda_{\nu l}$ in \eqref{eq:lambda} is 
of the form 
\begin{align}
\lambda_{\nu l}=&(-1)^{J-J^\prime+1}
\frac{\img^{-l}\nu^2}{\sqrt{l(l+1)}}
\frac{\sqrt{4\pi(2l+1)}}{\sqrt{2J+1}}  
\cgc{J^\prime}{l}{J}{M^\prime}{\rho}{M}
\label{eq:lambda2}
\end{align}
and the multipole moment is given by 
\begin{equation}
Q^l=[rj_l(\nu r)]^\prime C^l+O(\nu/c)
\label{eq:Ql2}
\end{equation}
where the prime denotes the derivative with respect to $r$.
Since we still have $\omega\ll c^2$, we do not give an
explicit representation for the term $O(\nu/c)$, which we 
omit in what follows.

Subsequently, the radial integral \eqref{eq:rad-1} is replaced by 
\begin{equation}
R_{\nu l}(\alpha_1,\alpha_2)=
\int_0^\infty \ol{R_{\alpha_1}(r)}
[rj_l(\nu r)]^\prime
R_{\alpha_2}(r)r^2\dd r\,.
\end{equation}
One verifies that $a^E_{\nu l}$ in \eqref{eq:aEM1}, with 
$\lambda_{\nu l}$ and $Q^l$ as in \eqref{eq:lambda2} and 
\eqref{eq:Ql2}, approaches $a^E_{\nu l}$,
with $\lambda_{\nu l}$ and $Q^l$ as in \eqref{eq:lambda} and 
\eqref{eq:QlMl}, as $\nu\searrow0$.

As an example, let us consider the Rydberg transition 
in Ag-like W$^{27+}$ ion \cite[Tab.~6]{Safr11}
\[4d^9 4f (LS) 9f \,^2\!G_{9/2}-4d^{10} 4f \,^2\!F_{7/2}\]
of wavelength $13.2$ ($\textrm{\AA}$), \ie 
$\nu\approx1.58417$. Let $(LS)=\!\!\,^1\!P$; then 
\begin{align*}
Q_{\nu 1}=&
\frac{3}{7}\sqrt{\frac{19}{2}}
R_{\nu1}(4d,9f)\,, 
\\ 
Q_{\nu 3}=&
\frac{1}{7}\sqrt{\frac{418}{21}}
R_{\nu3}(4d,9f)\,, 
\\ 
Q_{\nu 5}=&
\frac{1}{7}\sqrt{\frac{247}{66}}
R_{\nu5}(4d,9f)\,.
\end{align*}
For $M=-M^\prime=1/2$, the matrix element \eqref{eq:eddss} now takes 
the value (in cm$^{-1}$)
\begin{align*}
&\braket{1s,1/2\vert (H^{ex}_{AME})^{E\,\prime\prime}_\nu \vert 
3d_-,1/2} 
\\ 
&\qquad \approx 
-1.47049\cdot10^{-14}\,\img\sin(\sigma_1-\sigma_3) 
\\ 
&\qquad \quad +
8.04766\cdot10^{-27}\,\img\sin(\sigma_3-\sigma_5)
\end{align*}
while the matrix element \eqref{eq:eddss0} is given by 
\[\braket{1s,\pm1/2\vert (H^{ex}_{AME})^{E\,\prime}_\nu \vert 
1s,\pm1/2}
\approx
\pm9.33947\cdot10^{-7}\,.\]
Since the expansion in powers of $\nu$ no longer makes sense,
and the analytical representation of the matrix elements is 
rather complicated and veiled, we show only numerical values. 

As expected, the eigenvalues of the 
corresponding matrices are much larger compared to those for the previous 
transitions. Yet the values are much smaller compared to 
the classical $O(c^{-2})$ corrections. On the other hand, the 
matrix element \eqref{eq:eddss00}, which now equals  
$\pm10.9953\sin(\omega t+\sigma_1)$ (cm$^{-1}$), allows one to 
expect a sizeable contribution to energy levels.
\section{Concluding remarks}
We present the AME coupling term as the summable series of 
irreducible tensor operators so that the whole information about 
the radiation source is contained in the coefficients of expansion. 
The form is convenient both for applying directly the Wigner--Eckart 
theorem and for changing the radiation source, depending on ones 
interest, without affecting the structure of tensor operator.

We give a rough estimation of the order 
of magnitude of the corresponding matrix elements by considering 
electronic transitions in atomic systems. The examples indicate that 
the most valuable contribution to energy levels comes from the 
time-dependent intrinsic part of the AME coupling, provided that 
the radiation source is described by electronic transitions at 
short wavelengths and that the charge number is large enough.

A rigorous investigation of the eigenvalue problem for the 
time-dependent Pauli operator \cite{Bou08} perturbed by the 
time-dependent part of AME coupling still needs to be done. 
From the analytical point of view it would also be of interest 
to investigate the change of the AME coupling when modifying the 
vector potential by using gauge transformations.

\end{document}